\documentclass[runningheads]{llncs}
\usepackage[T1]{fontenc}
\usepackage{graphicx}
\usepackage{adjustbox}
\usepackage{bm}
\usepackage{booktabs}
\usepackage{multirow}
\usepackage{multicol}
\usepackage{color, colortbl}
\usepackage{pifont}
\usepackage{amsmath}
\usepackage{amsfonts}
\usepackage{amssymb}
\usepackage{algorithm}
\usepackage{algpseudocode}
\usepackage{listings}
\usepackage[colorlinks=true, linkcolor=blue, urlcolor=blue, citecolor=blue, anchorcolor=blue]{hyperref}
\usepackage{xspace}
\usepackage{lineno}
\usepackage{graphicx}

\usepackage[capitalize]{cleveref}
\Crefname{section}{Section}{Sections}
\crefname{section}{Sec.}{Secs.}
\Crefname{align}{Equation}{Equations}
\crefname{align}{Eq.}{Eqs.}
\Crefname{equation}{Equation}{Equations}
\crefname{equation}{Eq.}{Eqs.}
\Crefname{figure}{Figure}{Figures}
\crefname{figure}{Fig.}{Figs.}
\Crefname{table}{Table}{Tables}
\crefname{table}{Tab.}{Tabs.}

\newcommand{\norm}[1]{\left\lVert#1\right\rVert}

\newcommand{\nameCOLOR}[1]{\textcolor{black}{#1}} 
\newcommand{\model}{\nameCOLOR{\mbox{CryoSAM}}\xspace}

\definecolor{Gray}{gray}{0.9}
\definecolor{Celadon}{rgb}{0.67, 0.88, 0.69}
\definecolor{Cream}{rgb}{1.0, 0.99, 0.82}

\DeclareMathOperator{\argmax}{argmax}

\newcommand{\Ec}{\mathcal{E}}

\newcommand{\Rb}{\mathbb{R}}

\newcommand{\Cv}{\mathbf{C}}

\newcommand{\Fv}{\mathbf{F}}

\newcommand{\Iv}{\mathbf{I}}

\newcommand{\Kv}{\mathbf{K}}

\newcommand{\Mv}{\mathbf{M}}

\newcommand{\Pv}{\mathbf{P}}
\newcommand{\Qv}{\mathbf{Q}}

\newcommand{\Sv}{\mathbf{S}}

\newcommand{\Xv}{\mathbf{X}}
\newcommand{\Yv}{\mathbf{Y}}
\newcommand{\Zv}{\mathbf{Z}}

\begin{document}
\title{Training-free CryoET Tomogram Segmentation}
\titlerunning{\model}
%
\author{
  Yizhou Zhao$^1$
  \and
  Hengwei Bian$^1$
  \and
  Michael Mu$^1$
  \and
  Mostofa R. Uddin$^1$
  \and \\
  Zhenyang Li$^2$
  \and
  Xiang Li$^1$
  \and
  Tianyang Wang$^2$
  \and
  Min Xu$^1\thanks{Corresponding author.}$
}
\authorrunning{Y. Zhao et al.}
%
\institute{Carnegie Mellon University, Pittsburgh PA 15213, USA \and University of Alabama at Birmingham, Birmingham AL 35294, USA}
\maketitle              
\begin{abstract}
Cryogenic Electron Tomography (CryoET) is a useful imaging technology in structural biology that is hindered by its need for manual annotations, especially in particle picking. Recent works have endeavored to remedy this issue with few-shot learning or contrastive learning techniques. However, supervised training is still inevitable for them. We instead choose to leverage the power of existing 2D foundation models and present a novel, training-free framework, \model. In addition to prompt-based single-particle instance segmentation, our approach can automatically search for similar features, facilitating full tomogram semantic segmentation with only one prompt. \model is composed of two major parts: 1) a prompt-based 3D segmentation system that uses prompts to complete single-particle instance segmentation recursively with Cross-Plane Self-Prompting, and 2) a Hierarchical Feature Matching mechanism that efficiently matches relevant features with extracted tomogram features. They collaborate to enable the segmentation of all particles of one category with just one particle-specific prompt. Our experiments show that \model outperforms existing works by a significant margin and requires even fewer annotations in particle picking. Further visualizations demonstrate its ability when dealing with full tomogram segmentation for various subcellular structures. Our code is available at: \url{https://github.com/xulabs/aitom}.
\end{abstract}
\section{Introduction}
\label{sec:intro}
The advancement of Cryogenic Electron Tomography (CryoET) makes it possible to capture macromolecular structures with native conformations at nanometer resolution \cite{doerr2017cryo}. In a typical CryoET pipeline, researchers prepare frozen-hydrated samples and expose them to electron beams for imaging. The sample is incrementally tilted, allowing for the collection of multi-view images, i.e., tilt-series. These images can be used for 3D reconstruction, resulting in a 3D density map, the tomogram. Further investigation requires particle picking to accurately localize and segment sub-cellular structures. To this end, most existing methods~\cite{zhou2023machine,de2023convolutional,wagner2019sphire,zhang2019advances7,gubins2020shrec8,hao2022vp12,moebel2021deep13,liu2024deepetpicker} resort to supervised training or template matching~\cite{frangakis2002identification9}, necessitating a large amount of laborious annotation. Some recent works propose to adopt few-shot learning~\cite{zhou2021one} or contrastive learning~\cite{huang2022accurate} techniques to ameliorate this issue. However, currently, there is still a need to train on several known categories or at least 20-50 annotations.

Looking out of the CryoET domain, recent years have witnessed a proliferation of general-purpose segmentation models. With the ability to condition on various types of inputs and accomplish different downstream segmentation tasks \cite{li2024qdformer,li2023towards,li2024paintseg,li2023robust,li2022hybrid}, SAM~\cite{kirillov2023segment} and SEEM~\cite{zou2024segment} have demonstrated a diverse range of capabilities. Furthermore, in the three-dimensional world, SA3D~\cite{cen2024segment} and LERF~\cite{kerr2023lerf} extend the ability of the implicit 3D representation NeRF~\cite{mildenhall2021nerf} with prompt-based segmentation and visual grounding. This progress inspires us to explore segmenting CryoET tomograms with general-domain foundation models. However, there are several obstacles. While we see a tremendous number of 2D foundation models, their counterparts for 3D are relatively scarce, e.g., a general volumetric segmentation model is still absent. Hence, bridging general-domain foundation models to CryoET analysis is not trivial. In addition, general-purpose segmentation models~\cite{kirillov2023segment,cen2024segment} are commonly instance-specific while semantic-agnostic. This limits their direct application to semantic-specific particle picking, which requires picking all particles of a category simultaneously.

To overcome these challenges, we present \model, a training-free approach for prompt-based CryoET tomogram segmentation. Our method introduces a prompt-based 3D segmentation pipeline, bridging the gap between 2D segmentation models and 3D volumetric segmentation. Our intuition is that the silhouettes of a particle are similar in adjacent tomogram slices. Hence, we can segment its 3D structure layer after layer by refining the segmentation mask from the previous plane. Formally, we achieve this by employing a Cross-Plane Self-Prompting mechanism, which recursively propagates and refines segmentation masks along one direction by prompting SAM~\cite{kirillov2023segment} with segmentation results from preceding planes. This allows us to segment one particle instance with a single prompt. To further segment all particles of a specific category comprehensively, we introduce a Hierarchical Feature Matching strategy for efficient instance-level feature matching. This approach eliminates the need for predefined templates~\cite{cen2024segment,wu2019template} and the extraction of subtomograms~\cite{zeng2023high}. Using the mean feature of prompted particles as the query, it filters out regions dissimilar to the query in a coarse-to-fine manner. After filtering, it proposes point prompts in a relatively low resolution and relies on the prompt-based 3D segmentation pipeline to achieve final segmentation results. These designs enable semantic segmentation over a full CryoET tomogram with a single prompt.

Our contributions can be summed up as follows:
\begin{itemize}
    \item We present a novel, training-free framework, \model, that takes a full CryoET tomogram and a set of user prompts as input and segments the prompted particle and all particles of the same category. This contrasts with current methods that require supervised training~\cite{zhou2023machine,de2023convolutional,huang2022accurate,zhou2021one}.
    \item We introduce Cross-Plane Self-Prompting, which enables 3D volumetric segmentation with 2D foundation models, significantly reducing the labor cost of annotation by leveraging its prompt-based nature.
    \item We propose a Hierarchical Feature Matching strategy to match instance-level particle features. It cuts down the runtime by 95\% compared with naive feature matching, being more efficient and convenient to use.
\end{itemize}
\begin{figure}[t]
    \begin{center}
        \includegraphics[width=\linewidth]{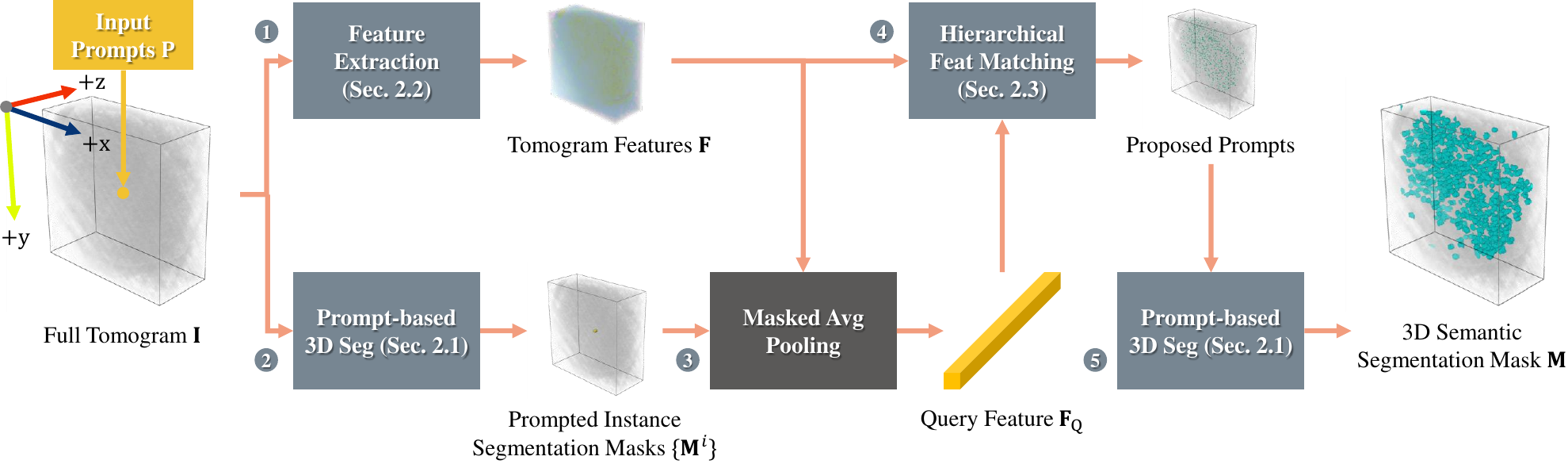}
        \caption{\textbf{Framework overview.} \ding{182}: We extract per-slice 2D features for three views (z, y, and x) from CryoET tomogram $\Iv$ and concatenate them as $\Fv$. \ding{183}: After segmenting the particle(s) prompted by $\Pv$ with instance segmentation mask(s), \ding{184}: we average pool the masked features to get query feature $\Fv_Q$. \ding{185}: To efficiently propose prompts for further segmentation, we match $\Fv_Q$ with $\Fv$ using Hierarchical Feature Matching. \ding{186}: Finally, we adopt prompt-based 3D segmentation for semantic segmentation results $\Mv$.}
        \label{fig:pipeline}
    \end{center}
\end{figure}

\section{Method}
Given a volumetric CryoET tomogram $\mathbf{I} \in \mathbb{R}^{D \times H \times W}$ and $N$ point prompts $\mathbf{P} \in \Rb^{N \times 3}$ denoting a set of single-category particles, our goal is to segment all particles of the same category as the prompted ones. This process predicts a 3D semantic segmentation mask $\Mv\in\{0,1\}^{D \times H \times W}$, with the overall pipeline depicted in~\cref{fig:pipeline}. $D$, $H$ and $W$ denote depth, height, and width respectively.

\subsection{Prompt-based 3D Segmentation}
\label{sub:prompt}
\begin{figure}[t]
    \begin{center}
        \includegraphics[width=\linewidth]{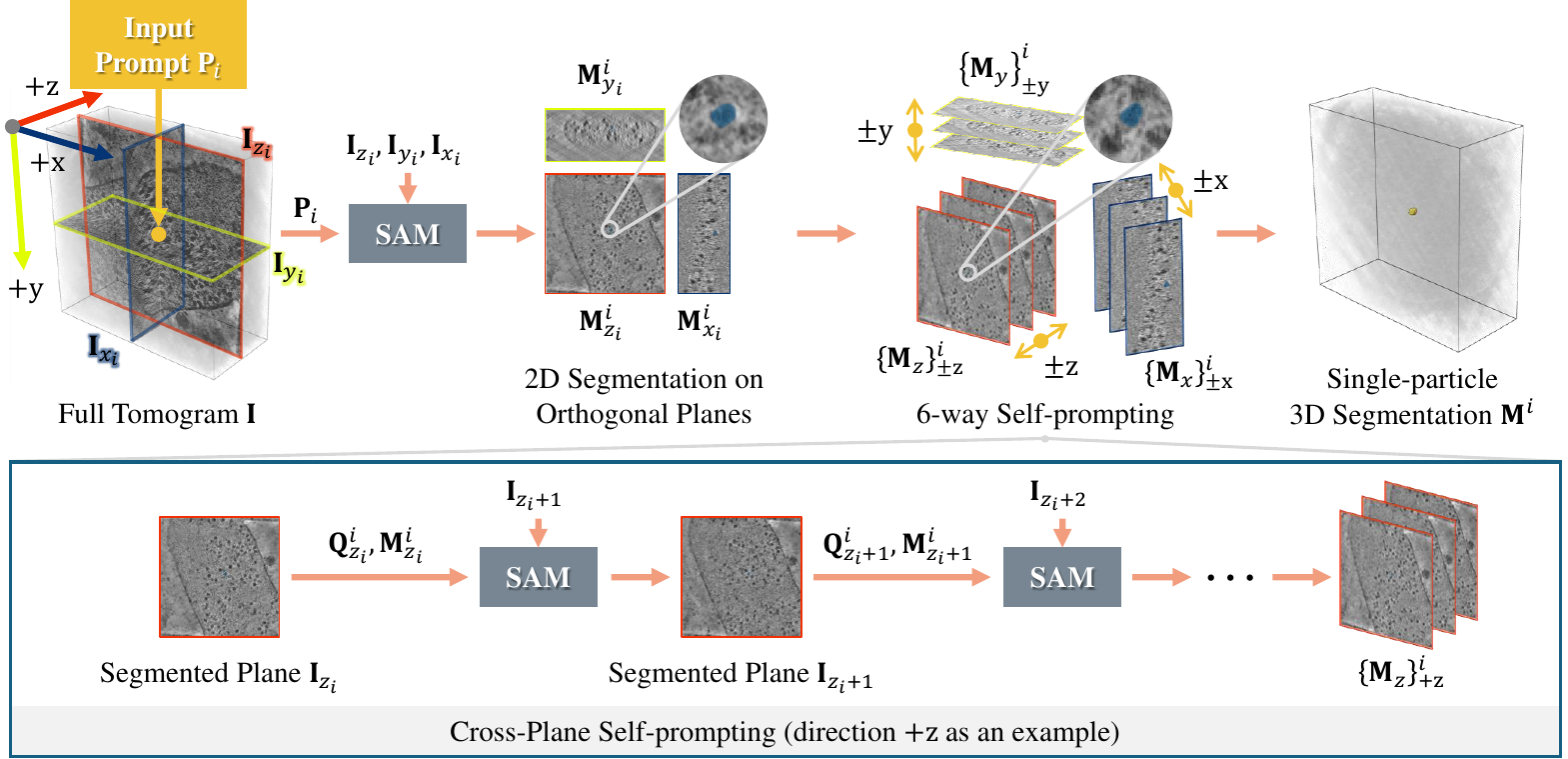}
        \caption{\textbf{The pipeline of prompt-based 3D segmentation.} After segmenting the orthogonal planes intersect at the point prompt $\Pv_i$, we iteratively execute Cross-Plane Self-Prompting until we get the complete mask of the particle.}
        \label{fig:prompt}
    \end{center}
\end{figure}

We propose Cross-Plane Self-Prompting, a mechanism that can propagate segmentation masks along the $\pm \text{z},\pm \text{y},\pm \text{x}$ axes, to approach prompt-based 3D segmentation, as illustrated in~\cref{fig:prompt}. The intuition is that the segmentation mask of one particle should be similar for neighboring slices. 
Hence, we can prompt SAM~\cite{kirillov2023segment} with the segmentation results from the previous plane to get subsequent results. Formally, we take as input a single point prompt $\Pv_i=[z_i,y_i,x_i]$ and the three orthogonal planes intersecting at this point, namely, the YX-plane $\Iv_{z_i}$, the ZX-plane $\Iv_{y_i}$, and the ZY-plane $\Iv_{x_i}$. Then, we employ SAM to obtain their 2D segmentation results, with the YX-plane as an example
\begin{align}
    (\Cv_{z_i}^{i}, \Mv_{z_i}^{i}) &= \text{SAM}\left[\Iv_{z_i}|(x_i,y_i)\right],\;
    \Qv_{z_i}^{i} = \argmax_{x,y}(\Cv_{z_i}^{i}),
\end{align}
where $\Cv_*^{i}$ are the predicted confidence scores, $\Mv_*^{i}$ are the predicted segmentation masks, and $\Qv_*^{i}$ are the coordinates with the highest confidence scores. We use superscript $^i$ to represent the index of the initial point prompt. Then for each direction in $\{\pm \text{z},\pm \text{y},\pm \text{x}\}$, we prompt the next tomogram slice with $\Qv_*^{i}$ and $\Mv_*^{i}$ from the previous plane, for which we term Cross-Plane Self-Prompting. Taking the $+\text{z}$ direction as an example which starts from $z=z_i$, we have
\begin{align}
    (\Cv_{z+1}^{i}, \Mv_{z+1}^{i}) &= \text{SAM}\left[\Iv_{z}|\Qv_{z}^{i},\Mv_{z}^{i}\right],\;
    \Qv_{z+1}^{i} = \argmax_{x,y}(\Cv_{z+1}^{i}).
\end{align}
Here, we benefit from SAM's versatility, which allows it to take both point and mask prompts as inputs. This recursive process continues until the intersection over union (IoU) of the segmentation masks in two adjacent slices drops below a threshold $\tau_\text{IoU}$, which suggests that prompting the current plane will not get a result consistent with previous ones. After getting the segmentation masks $\{\Mv_z\}_{\pm \text{z}}^{i},\{\Mv_y\}_{\pm \text{y}}^{i},\{\Mv_x\}_{\pm \text{x}}^{i}$ for all 6 directions, we aggregate a union of all segmentation masks in 3D, i.e.,
$
    \Mv^{i} = \left\{\Mv_z\right\}_{\pm \text{z}}^{i}\cup\left\{\Mv_y\right\}_{\pm \text{y}}^{i}\cup\left\{\Mv_x\right\}_{\pm \text{x}}^{i}.
$

\begin{figure}[t]
    \begin{center}
        \includegraphics[width=\linewidth]{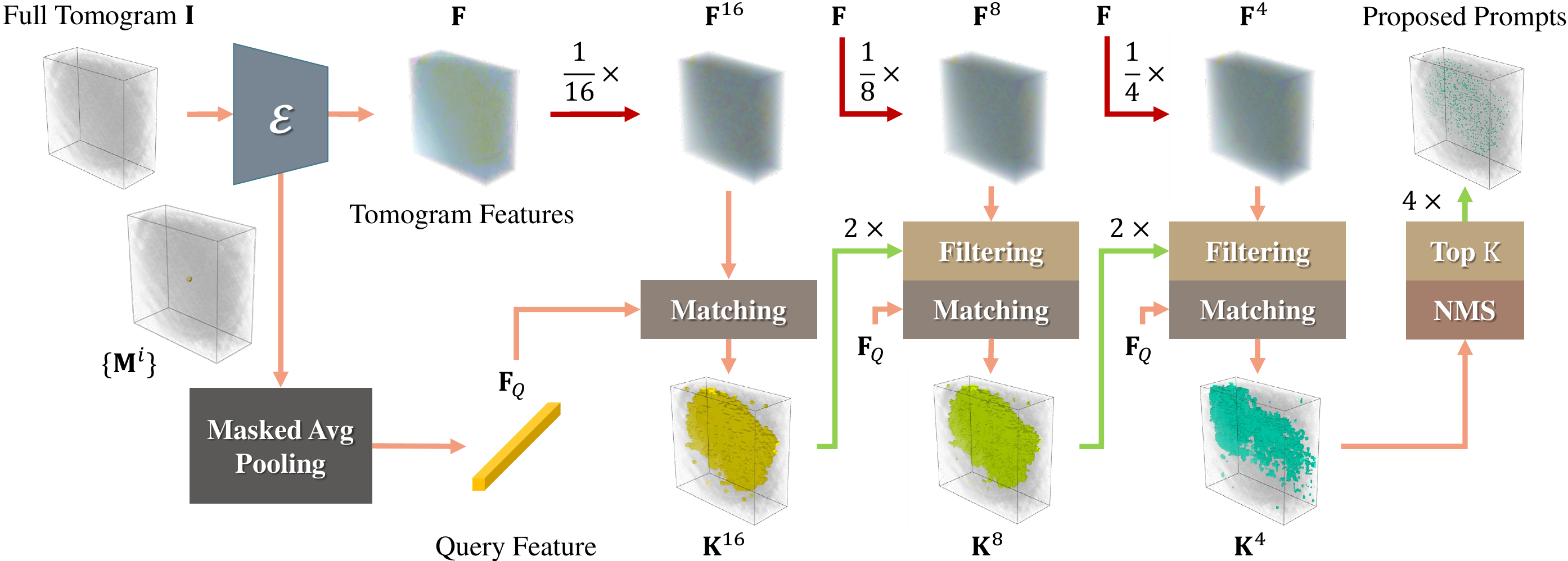}
        \caption{\textbf{The pipeline of Hierarchical Feature Matching.} We average the tomogram features in the instance segmentation masks to obtain a query feature $\Fv_Q$. Then we downsample $\Fv$ into several coarse ones and match them with $\Fv_Q$ in a coarse-to-fine manner. After the last matching stage, we apply NMS and gather coordinates with top $K$ similarities as prompts to derive final semantic segmentation results.}
        \label{fig:match}
    \end{center}
\end{figure}

\subsection{Feature Extraction}
\label{sub:extract}
We rely on an off-the-shelf image encoder $\Ec$ to extract 2D features from tomogram slices $\{\Iv_z\}_{z=1}^D,\{\Iv_y\}_{y=1}^H,\{\Iv_x\}_{x=1}^W$. For each view z, y, and x, we obtain $\Zv^{\Ec}=\{\Ec(\Iv_z)\}_{z=1}^D\in\Rb^{D \times h \times w \times C}$, $\Yv^{\Ec}=\{\Ec(\Iv_y)\}_{y=1}^H\in\Rb^{d \times H \times w \times C}$, and $\Xv^{\Ec}=\{\Ec(\Iv_x)\}_{x=1}^W\in\Rb^{d \times h \times W \times C}$, where the lowercase $d,h,w$ are feature resolutions in the latent space. Then we bilinear upsample them to get $\Zv,\Yv,\Xv\in\Rb^{D \times H \times W \times C}$, and aggregate them with a concatenation
\begin{align}
    \Fv=\{\Fv_{zyx}\}_{z=1,y=1,x=1}^{D,H,W}=\left[\Zv, \Yv, \Xv\right]\in\Rb^{D \times H \times W \times 3C},
\end{align}
where $\Fv_{zyx}$ is a feature vector in $\Fv$ with coordinates $[z,y,x]$.

\subsection{Hierarchical Feature Matching}
\label{sub:match}
Shown in~\cref{fig:match}, Hierarchical Feature Matching aims to efficiently search for voxel regions with similar features as the query. For input point prompts $\Pv=\{\Pv_i\}\in\Rb^{N \times 3}$, we obtain an instance segmentation mask for each prompt through prompt-based 3D segmentation, resulting in $\{\Mv^{i}\}$. Then, we derive the query feature $\Fv_Q$ via masked average pooling (MAP)
\begin{align}
    \Fv_Q = \frac{\sum_i\sum_{zyx}\Mv_{zyx}^{i}\odot\Fv_{zyx}}{\sum_i\norm{\Mv^{i}}_0},
\end{align}
where $\odot$ is the Hadamard product with broadcasting and $\norm{\cdot}_0$ is the 0-norm indicating the number of non-zero voxels. This operation averages features masked by the instance segmentation masks to obtain a mean feature representing the prompted particles. While a brute-force approach can achieve voxel-precise feature matching between $\Fv_Q$ and $\Fv$, we empirically show this is neither efficient nor necessary. Instead, we propose to match $\Fv_Q$ with multi-resolution features in $\Fv$ in a coarse-to-fine manner, each time keeping only the most similar proportion. We begin with building a feature pyramid
\begin{align}
    \{\Fv^{r}\} = \left\{\left[\Zv^{r}, \Yv^{r}, \Xv^{r}\right]\right\},
\end{align}
where $r\in\{16,8,4\}$ is the downsampling ratio, and $\Fv^{r} \in \Rb^{\frac{D}{r} \times \frac{H}{r} \times \frac{W}{r} \times 3C}$. $\Zv^{r}\in\Rb^{\frac{D}{r} \times \frac{H}{r} \times \frac{W}{r} \times C}$ stands for an $r$ times downsampled version of $\Zv$, with similar definitions for $\Yv^{r}$ and $\Xv^{r}$.
Then from the lowest resolution of $\{\Fv^{r}\}$, we calculate its point-wise cosine similarity $\Sv^{r} = \{\Sv_{zyx}^{r}\}_{z=1,y=1,x=1}^{\frac{D}{r}, \frac{H}{r}, \frac{W}{r}}$ with query $\Fv_Q$
\begin{align}
    \Sv_{zyx}^{r}=\frac{\Fv_Q\cdot(\Fv_{zyx}^{r})^\top}{\norm{\Fv_Q}_2\cdot\norm{\Fv_{zyx}^{r}}_2}.
\end{align}
For the lowest resolution, we calculate the similarity for all $\frac{D}{r} \frac{H}{r} \frac{W}{r}$ features. Subsequently, we build a mask $\Kv^{r}=\Sv^{r} \geq \tau_\text{sim}$ that filters out regions with low similarity scores and propagates this mask to the next resolution with upsampling. This allows the next round of feature matching to be conducted only on the high-similarity features, thereby greatly reducing the computational complexity. After iterating through the whole downsampling ratio list, we apply non-maximum suppression (NMS) on the coordinates with their similarity scores and keep the top $K$ of them as point prompts. These prompts are then fed into the prompt-based 3D segmentation pipeline for semantic segmentation.

\begin{table}[t]
\caption{\textbf{Comparison results for particle picking on EMPIAR-10499~\cite{tegunov2021multi}.}}
\label{tab:sota}
\centering
\footnotesize
\setlength{\tabcolsep}{0.4em}
\adjustbox{width=\linewidth}{
\begin{tabular}{cc|cccc}
    \toprule
        Method & Annotation Ratio & Precision & Recall & F1 Score & Runtime (min) \\
    \midrule
        EMAN2~\cite{tang2007eman2} & - & 26.1 & 55.3 & 35.5 & 2-5 \\
        \midrule
        crYOLO~\cite{wagner2019sphire} & 100\% & 47.8 & 56.8 & 52.0 & 30-40 \\
    \midrule
        \multirow{5}{*}{Huang et al.~\cite{huang2022accurate}} & 5\% & 49.6 & 58.1 & 53.5 & \multirow{5}{*}{5-10} \\
        & 10\% & 50.1 & 58.2 & 53.8 & \\
        & 30\% & 55.9 & 60.3 & 58.0 & \\
        & 50\% & 53.0 & 65.1 & 58.4 & \\
        & 70\% & 54.9 & 66.7 & 60.2 & \\
    \midrule
        \rowcolor{Gray} & $<1\%$ (single prompt) & 53.1 & 55.3 & 54.2 &  \\
        \rowcolor{Gray} & 5\% & 57.8 & 74.3 & 65.0 &  \\
        \rowcolor{Gray} & 10\% & 58.2 & 75.1 & 65.5 &  \\
        \rowcolor{Gray} & 30\% & 58.1 & 75.4 & 65.6 &  \\
        \rowcolor{Gray} & 50\% & 58.0 & 75.3 & 65.5 &  \\
        \rowcolor{Gray}\multirow{-6}{*}{\model (Ours)} & 70\% & 58.5 & 79.4 & 67.4 & \multirow{-6}{*}{10-15} \\
    \bottomrule
\end{tabular}
}
\end{table}

\section{Experiment}
\subsection{Experimental Settings}
\paragraph{Datasets and evaluation metrics.}
Due to the scarcity of CryoET segmentation annotations, we mainly assess the quantitative performance of \model for particle picking. To this end, we utilize the EMPIAR-10499 dataset \cite{tegunov2021multi,iudin2016empiar}, which comprises 65 tilt-series of native M. pneumoniae cells with annotated ribosomes. We use the prediction from each proposed prompt as an instance segmentation mask to compare with other detection methods~\cite{huang2022accurate,tang2007eman2,wagner2019sphire} in terms of precision, recall, and F1 score. Results from all 65 tilt-series are averaged in our comparison results reported in~\cref{tab:sota}, while the first 20 are used in our ablation study. We do not calculate mean average precision (mAP) as our method does not output an explicit score for each segmentation mask.

\paragraph{Implementation details.} 
We use DINOv2~\cite{oquab2023dinov2} with a ViT-L/14~\cite{dosovitskiy2020image} backbone as the default 2D encoder of \model and SAM~\cite{kirillov2023segment} with ViT-H as our 2D segmentation model. The IoU threshold $\tau_\text{IoU}$ to determine the end of segmentation mask propagation and the similarity threshold $\tau_\text{sim}$ to filter out dissimilar regions in Hierarchical Feature Matching are both set to 0.5. Top $K=512$ coordinates in the final stage of Hierarchical Feature Matching are used as prompts for full tomogram semantic segmentation. In all experiments, we do not require any training for \model. We use a subset of all ground truth coordinates as input prompts. The annotation ratio in tables refers to the proportion of prompted particles to all particles in our scenario.

\subsection{Comparison Results}
\label{subsec:sota}
In~\cref{tab:sota}, \model demonstrates significant advancements in particle picking compared to three baselines under the same annotation ratio. It is noteworthy that our single-prompt result is better than the performance of Huang et al.~\cite{huang2022accurate} under 10\% annotation, which shows the annotation-efficient property of \model. Our performance also improves as the number of available prompts increases. This is probably because the averaged features are more robust with the addition of different particle instances in similarity-based matching.

\begin{table}[t]
\caption{\textbf{Ablation study for different feature extractors.}}
\label{tab:feature}
\centering
\footnotesize
\setlength{\tabcolsep}{0.4em}
\adjustbox{width=0.8\linewidth}{
\begin{tabular}{cc|ccc}
    \toprule
        2D Feature Extractor & Annotation Ratio & Precision & Recall & F1 Score \\
    \midrule
        \multirow{2}{*}{SAM~\cite{kirillov2023segment}} & $<1\%$ (single prompt) & 37.4 & 38.8 & 38.1 \\
        & 10\% & 44.1 & 60.0 & 50.8 \\
    \midrule
        \multirow{2}{*}{DINO~\cite{caron2021emerging}} & $<1\%$ (single prompt) & 56.3 & 52.8 & 54.5 \\
        & 10\% & 63.2 & 74.4 & 68.3 \\
    \midrule
        \multirow{2}{*}{DINOv2~\cite{oquab2023dinov2}} & $<1\%$ (single prompt) & 55.4 & 58.8 & 57.1 \\
        & 10\% & 59.8 & 80.1 & 68.5 \\
    \bottomrule
\end{tabular}
}
\end{table}

\begin{table}[t]
\caption{\textbf{Ablation study for different feature matching strategies.}}
\label{tab:match}
\centering
\footnotesize
\setlength{\tabcolsep}{0.4em}
\adjustbox{width=\linewidth}{
\begin{tabular}{cc|cccc}
    \toprule
        Feature Matching Strategy & Annotation Ratio & Precision & Recall & F1 Score & Runtime (min) \\
    \midrule
        \multirow{2}{*}{Naive} & $<1\%$ (single prompt) & 53.5 & 56.4 & 54.9 & \multirow{2}{*}{60-65} \\
        & 10\% & 60.8 & 80.7 & 69.4 &  \\
        \midrule
        \multirow{2}{*}{Hierarchical} & $<1\%$ (single prompt) & 55.4 & 58.8 & 57.1 & \multirow{2}{*}{10-15} \\
        & 10\% & 59.8 & 80.1 & 68.5 \\
    \bottomrule
\end{tabular}
}
\end{table}

\subsection{Ablation Study and Analysis}
\label{subsec:ablation}
\paragraph{Impact of feature extractors.}
We ablate the particle picking performance over different 2D feature extractors in~\cref{tab:feature}. Our results show that using DINO~\cite{caron2021emerging} and DINOv2~\cite{oquab2023dinov2} achieves significantly better results than using the SAM~\cite{kirillov2023segment} encoder. It follows that DINO and DINOv2 learn more discriminative features with self-supervised training, which is beneficial for accurate feature matching.

\paragraph{Impact of feature matching strategies.}
We evaluate the effectiveness of Hierarchical Feature Matching in~\cref{tab:match} by replacing it with naive feature matching that only computes voxel-wise similarity in the highest $DHW$ resolution. We see our hierarchical strategy retains a comparable performance while taking a notably shorter time to process. This reflects the robustness of our prompt-based 3D segmentation pipeline, which does not require the proposal to be voxel-precise.

\begin{figure}[t]
    \centering
    \includegraphics[width=\linewidth]{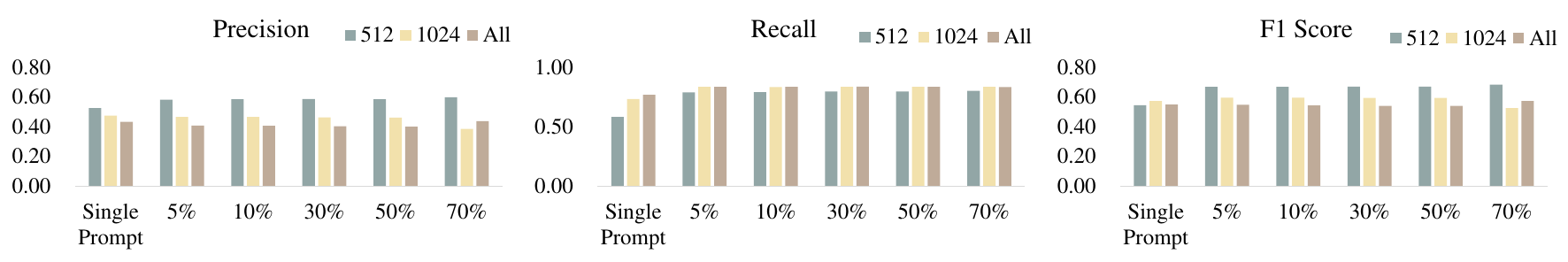}
    \caption{\textbf{Ablation study for the number of proposed prompts.} 512/1024/All: number of proposed prompts selected for prompt-based semantic segmentation. }
    \label{fig:top_k}
\end{figure}

\paragraph{Impact of the number of proposed prompts.}
In~\cref{fig:top_k}, we analyze the precision-recall trade-off by varying $K$. Generally, smaller values of $K$ result in lower recall and higher precision. We make our design choice to set $K=512$ by selecting the model with the best overall F1 score.

\begin{figure}[t]
    \begin{center}
        \includegraphics[width=\linewidth]{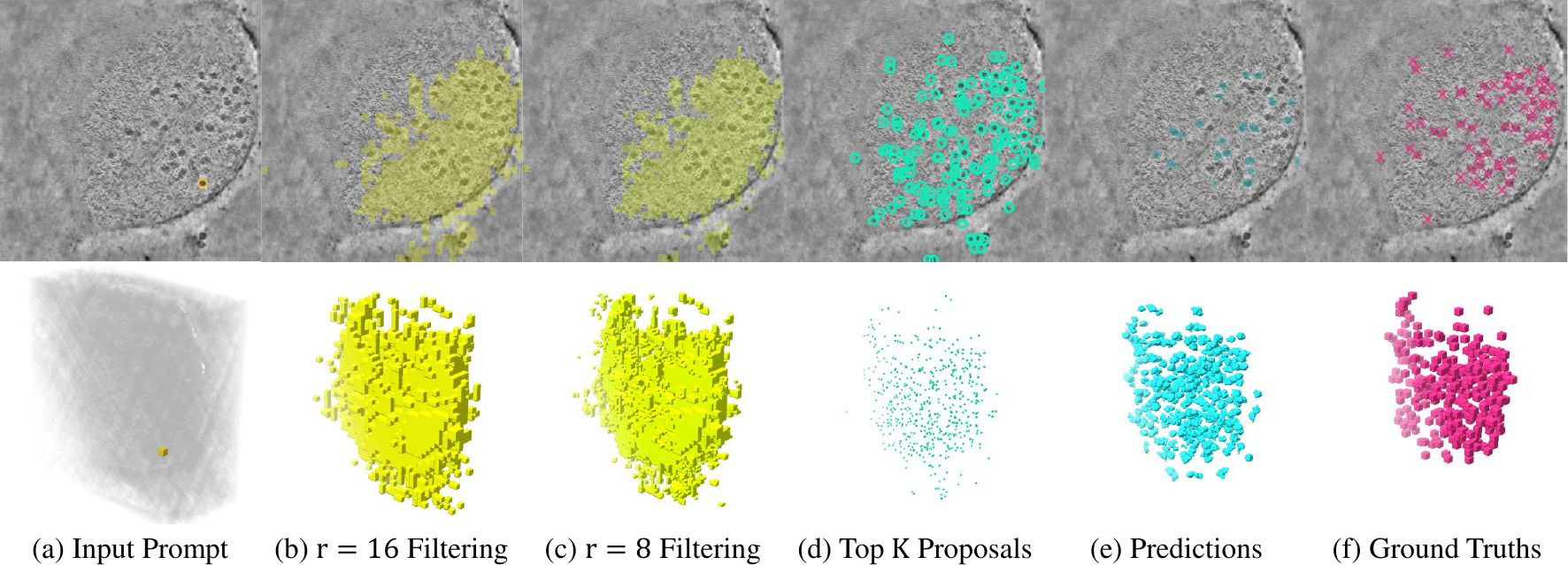}
        \caption{\textbf{Intermediate and final results of \model.} In (d) and (f), we show points with coordinates ranging from $z-20$ to $z+20$ for demonstration.}
        \label{fig:visual}
    \end{center}
\end{figure}

\paragraph{Qualitative analysis.}
We visualize the whole process of \model in~\cref{fig:visual}, which shows it can conduct 3D semantic segmentation with just a single point prompt. See the supplementary for more qualitative results and failure cases.
\section{Conclusion}
We present \model, a training-free framework that segments full CryoET tomograms with given prompts. It has two core innovations. First, the proposed Cross-Plane Self-Prompting mechanism bridges the gap between 2D segmentation foundation models and 3D volumetric segmentation. Second, we introduce Hierarchical Feature Matching, which is capable of efficient search for one category of particles. Combining both shows positive synergy in prompt-based full tomogram semantic segmentation, leading to SOTA results in particle picking.
\begin{credits}
\subsubsection{\ackname} This study was partially funded by U.S. NIH grants R01GM134020 and P41GM103712, NSF grants DBI-1949629, DBI-2238093, IIS-2007595, IIS-2211597, and MCB-2205148. Additionally, it received support from Oracle Cloud credits and resources provided by Oracle for Research, as well as computational resources from the AMD HPC Fund. MRU was supported by a fellowship from CMU CMLH.
\end{credits}
%
%
%
\clearpage
\bibliographystyle{splncs04}
\bibliography{bibliography}
\clearpage
\section*{Supplementary Material}

\begin{figure}[ht]
    \begin{center}
        \includegraphics[width=\linewidth]{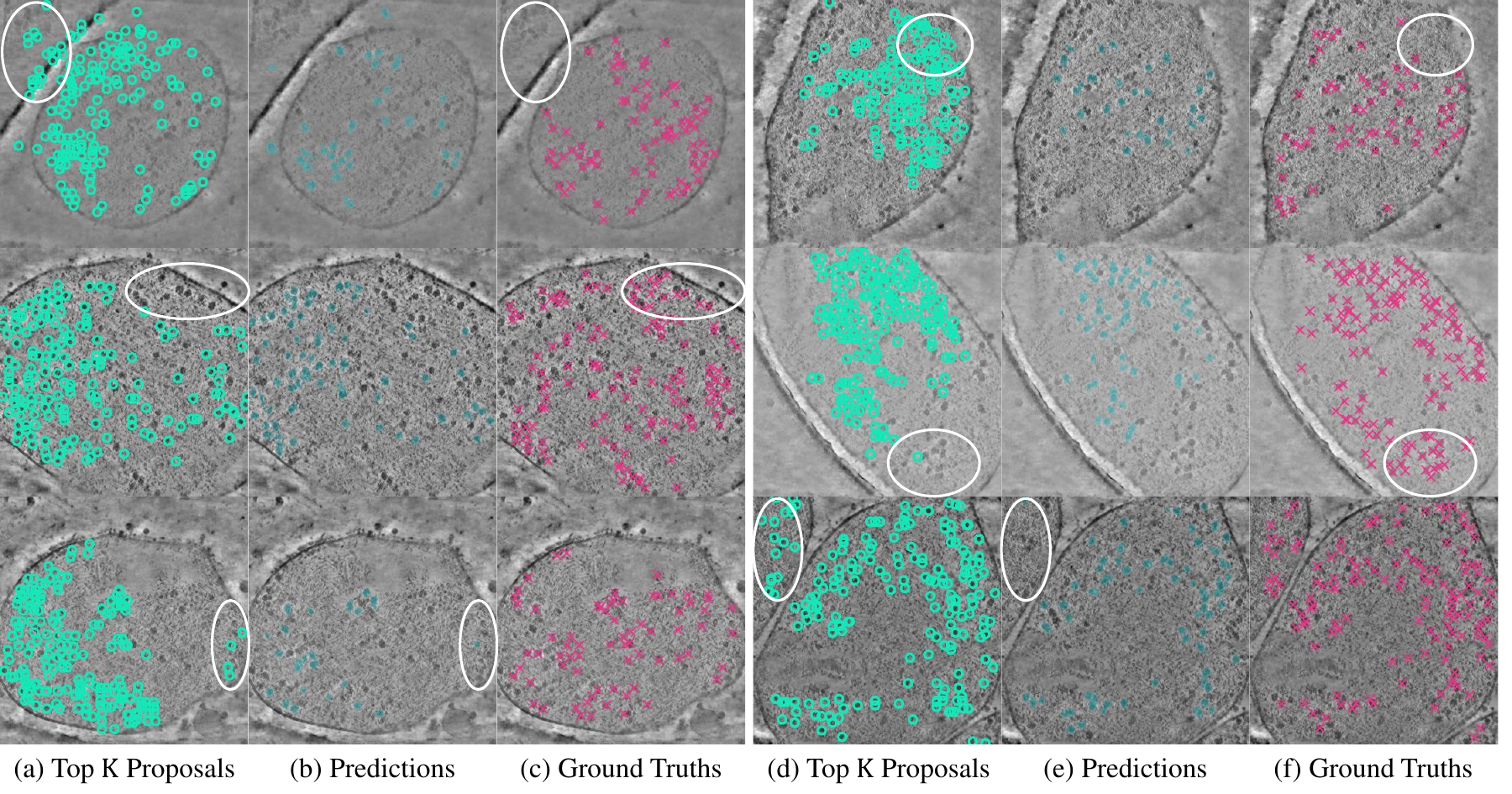}
        \caption{\textbf{Failure cases of \model.} 1st row: False Positive in proposed prompts. 2nd row: False Negative in proposed prompts. 3rd row: False Negative in final predictions.}
    \end{center}
\end{figure}

\begin{figure}[hb]
    \begin{center}
        \includegraphics[width=\linewidth]{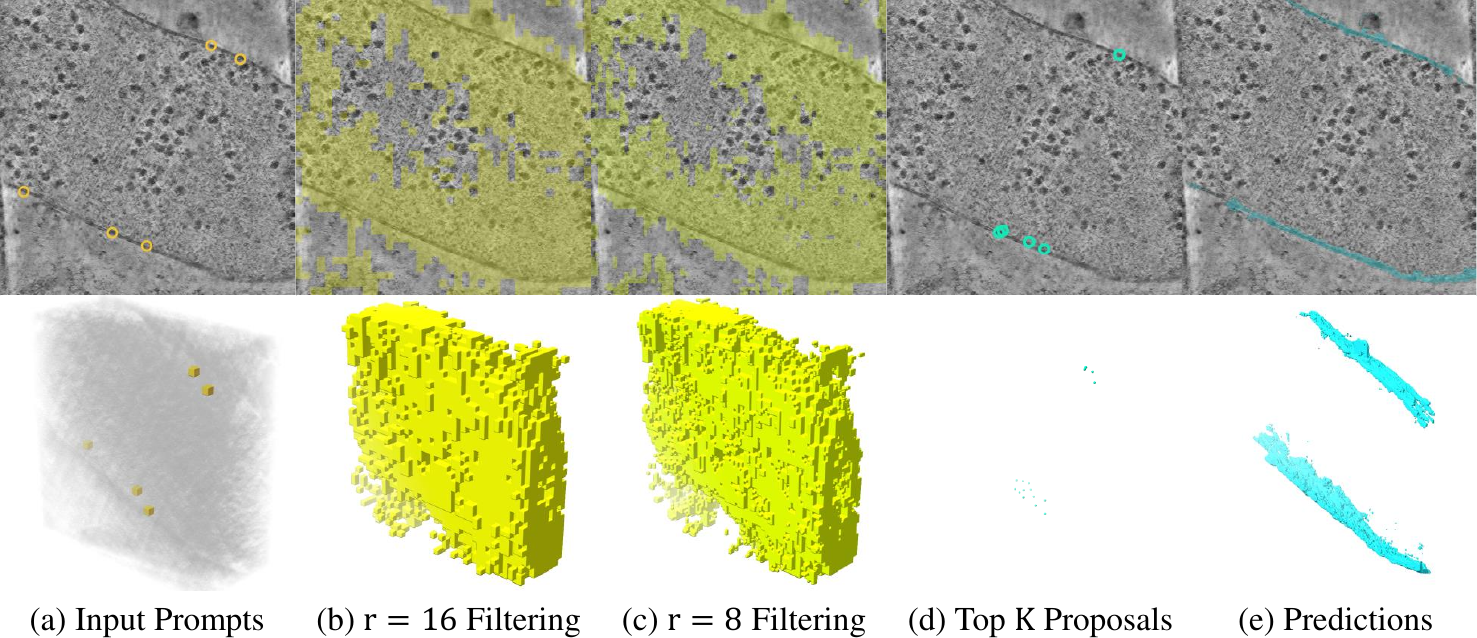}
        \caption{\textbf{Intermediate and final predictions of \model for membrane segmentation.} \model can segment membranes with sparse prompt inputs.}
    \end{center}
\vspace{-10.0em}
\end{figure}

\begin{figure}[ht]
    \begin{center}
        \includegraphics[width=\linewidth]{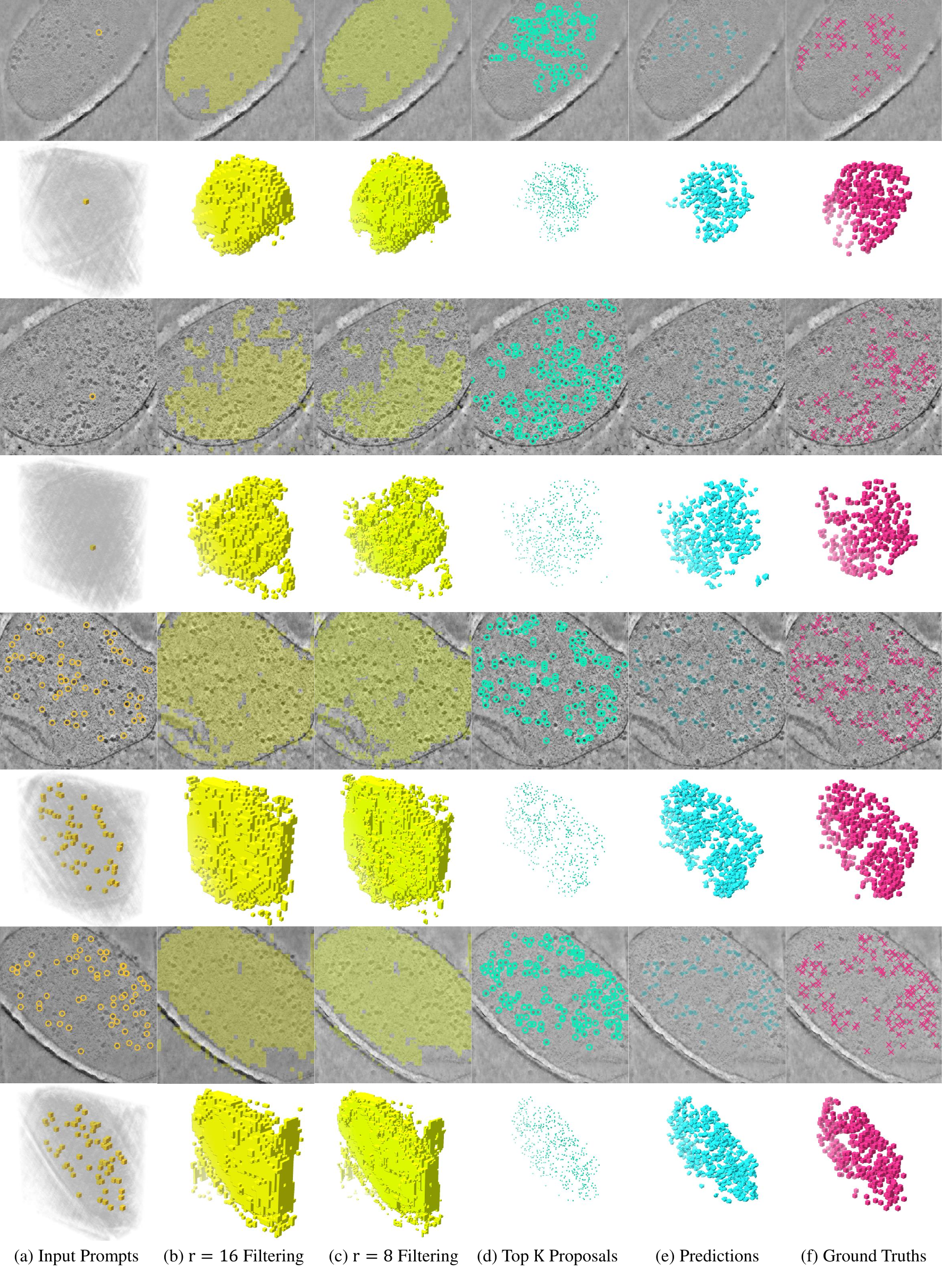}
        \caption{\textbf{Intermediate and final predictions of \model for particle picking.} We provide additional results for feeding \model with both single-point prompts and multiple-point prompts. In columns (a), (d), and (f), we show points with coordinates ranging from $z-20$ to $z+20$ for demonstration, where $z$ is the coordinate of the visualized tomogram slice.}
        \label{fig:success}
    \end{center}
\end{figure}
\end{document}